\documentclass[aps,prl,twocolumn,superscriptaddress,bibliography,nobibnotes]{revtex4-2}
\usepackage{amsfonts}
\usepackage{mathrsfs}
\usepackage{amsmath}
\usepackage{xcolor}
\usepackage{graphicx}
\usepackage{bm}
\usepackage{amssymb}
\usepackage{xspace}
\usepackage{epstopdf}
\usepackage{dcolumn}
\usepackage{longtable}
\usepackage{multirow}
\usepackage{float}
\usepackage{comment}
\usepackage{wasysym}
\usepackage{esint}
\usepackage[colorlinks=true, letterpaper=true, pdfstartview=FitV,  linkcolor=blue, citecolor=blue, urlcolor=blue]{hyperref}



\usepackage{babel}
\begin{document}
\title{Quantum-Metric-Induced Distinct Optoelectronic Responses in Triple $M$ Valleys}

\author{Yongliang Song}
\affiliation{Center for Quantum Transport and Thermal Energy Science, Institute of Physics Frontiers and Interdisciplinary Sciences, School of Physics and Technology, Nanjing Normal University, Nanjing 210023, China}

\author{Shuxian Sun}
\affiliation{Center for Quantum Transport and Thermal Energy Science, Institute of Physics Frontiers and Interdisciplinary Sciences, School of Physics and Technology, Nanjing Normal University, Nanjing 210023, China}

\author{Zhichao Zhou}
\affiliation{Center for Quantum Transport and Thermal Energy Science, Institute of Physics Frontiers and Interdisciplinary Sciences, School of Physics and Technology, Nanjing Normal University, Nanjing 210023, China}

\author{Xiao Li}
\email{lixiao@njnu.edu.cn}
\affiliation{Center for Quantum Transport and Thermal Energy Science, Institute of Physics Frontiers and Interdisciplinary Sciences, School of Physics and Technology, Nanjing Normal University, Nanjing 210023, China}

\begin{abstract}  
While $K_{\pm}$ valleys with contrasting optoelectronic properties have been the main focus of valleytronics, the lack of reliable manipulation methods for $M$ valleys leaves numerous $M$-valley materials untapped.
Here, we discover that three flavors of $M$ valleys become distinguishable via quantum-metric-induced optoelectronic responses.
The tight-binding calculations demonstrate that by tuning the electric-field orientation, opposite nonlinear Hall currents from any pair of $M$ valleys can be chosen to realize the nonlinear valley Hall effect.
Moreover, linearly polarized light allows for preferential excitation of individual $M$ valleys, driving photoexcited valley-polarized Hall transport.
Using first-principles calculations, we further validate the quantum-metric-induced optoelectronic responses in antiferromagnetic $\mathrm{MoI}_3$ bilayers, which exhibit a significant nonlinear valley Hall conductivity amenable to transport measurements.
Our findings extend the scope of valleytronics from conventional binary $K_{\pm}$ valleys to triple $M$ valleys and pave the way for designing advanced optoelectronic devices with multiple quantum states of high tunability.
  
\end{abstract}

\maketitle
Valleys, as local extrema in the band structure, constitute an emergent electronic degree of freedom in addition to charge and spin, with enormous applications in information storage and manipulation~\cite{Rycerz2007,Xiao2007,Schaibley2016,Vitale2018,Seyler2026}.
Previous research on valley physics has largely focused on inequivalent $K_{\pm}$ valleys in a hexagonal lattice.
At $K_{\pm}$ valleys, carriers are selectively excited by circularly polarized light, and valley-contrasting Berry curvatures generate opposite anomalous Hall currents, giving rise to the valley Hall effect~\cite{Yao2008,Xiao2012,Cao2012,Zeng2012,Mak2012,Mak2014}.
Besides $K_{\pm}$ valleys, numerous materials host other multiple inequivalent valleys, such as $M$ valleys in a hexagonal lattice~\cite{Sattar2022,Li2026CoSe,Chen2022,Yu2024,Calugaru2025,Lei2025}.
Different from $K_{\pm}$ valleys that come in two selectable states, $M$ valleys comprise three inequivalent counterparts and would in principle offer greater versatility for valley control.
However, the selective manifestation of different $M$ valleys represents an outstanding challenge.
$M$ valleys typically exhibit vanishing circular dichroism and Berry curvature~\cite{Cao2012,Liu2013}, which greatly limits applications of $M$ valleys in valleytronics.
Therefore, it is highly desirable to explore novel mechanisms for manipulating $M$ valleys, thereby advancing and enriching valleytronics based on distinct valleys.

Electrical nonlinear transport has recently emerged as a rapidly growing topic in condensed matter physics~\cite{Gao2014,Sodemann2015,Du2018,Wang2021,Liu2021,Du2021,Fang2024,Suarez2025,Xiang2026}, and possibly provides a viable avenue for exploring valley-contrasting physics of $M$ valleys.
The nonlinear Hall effect denotes a higher-order transverse current response to an applied electric field, and arises from, e.g., the Berry-connection polarizability (BCP) associated with the quantum metric (QM)~\cite{Wang2021,Liu2021} or the Berry-curvature dipole~\cite{Sodemann2015,Du2018}. 
Given that the nonlinear Hall effect requires the breaking of threefold rotational symmetry, there is no nonlinear current from $K_{\pm}$ valleys unless an external stimulus such as uniaxial strain is introduced~\cite{Das2024,Zhou2025}. 
In contrast, each $M$ valley inherently breaks the symmetry, potentially enabling a finite nonlinear Hall current and thus endowing $M$ valleys with an intrinsic advantage over $K_{\pm}$ valleys in the nonlinear transport regime. 
However, relevant studies remain absent, and it is still unclear what nonlinear-transport-based manipulation schemes, differing from those of $K_{\pm}$ valleys, may emerge for $M$ valleys.
Besides, a QM-related optical selection rule has been proposed~\cite{Li2026Optical}, where the mirror symmetry at a given $M$ valley dictates that valley carriers can selectively couple to linearly polarized light with polarization parallel or perpendicular to the mirror plane.
Although this mechanism has been established in a hexagonal lattice, it remains an open question how to preferentially excite all three $M$ valleys.

In this work, we investigate the QM-induced optoelectronic effect from triple $M$ valleys. 
A tight-binding model of a $\mathcal{PT}$-symmetric antiferromagnetic honeycomb lattice demonstrates that each $M$ valley indeed generates a nonvanishing nonlinear Hall conductivity.
By rotating the direction of the applied electric field, opposite nonlinear Hall currents from any two $M$ valleys can be selectively accessed, realizing the nonlinear valley Hall (NVH) effect with the full utilization of three flavors of $M$ valleys, which contrasts sharply with the electric-field-independent valley Hall effect from $K_{\pm}$ valleys. 
Furthermore, $M$ valleys are found to be preferentially excited by linearly polarized light, leading to a photoexcited valley-polarized Hall effect with a net anomalous Hall current.
The QM-induced optoelectronic responses are then verified in a realistic material, the $\mathrm{MoI}_3$ bilayer, using first-principles calculations. 
The bilayer possesses a significant NVH conductivity of 27.1 $(e^3/\hbar) \, \text{\AA} \, \text{eV}^{-1}$, which falls within the experimentally detectable range~\cite{Godinho2018,Wang2023,He2026}.
Our work utilizes the QM-induced optoelectronic effect to render triple $M$ valleys manipulable through a variety of approaches, thereby extending the pool of material candidates for valley physics and offering a practical route toward advanced multistate quantum devices based on nonlinear transport.

\textit{\color{blue} Tight-binding model.}
To investigate the valley-contrasting optoelectronic properties of triple $M$ valleys, we first construct a tight-binding model of a two-dimensional hexagonal antiferromagnet.
As a concrete platform, a centrosymmetric honeycomb lattice with two sublattices is considered, which is analogous to graphene. 
The lattice has a crystalline space group of $P6/mmm$ (No.~ 191). 
This space group contains the sixfold rotation $\mathcal{C}_{6z}$, the spatial inversion $\mathcal{P}$, and vertical mirrors.

\begin{figure}[b]
	\begin{centering}
		\includegraphics[width=8.4cm]{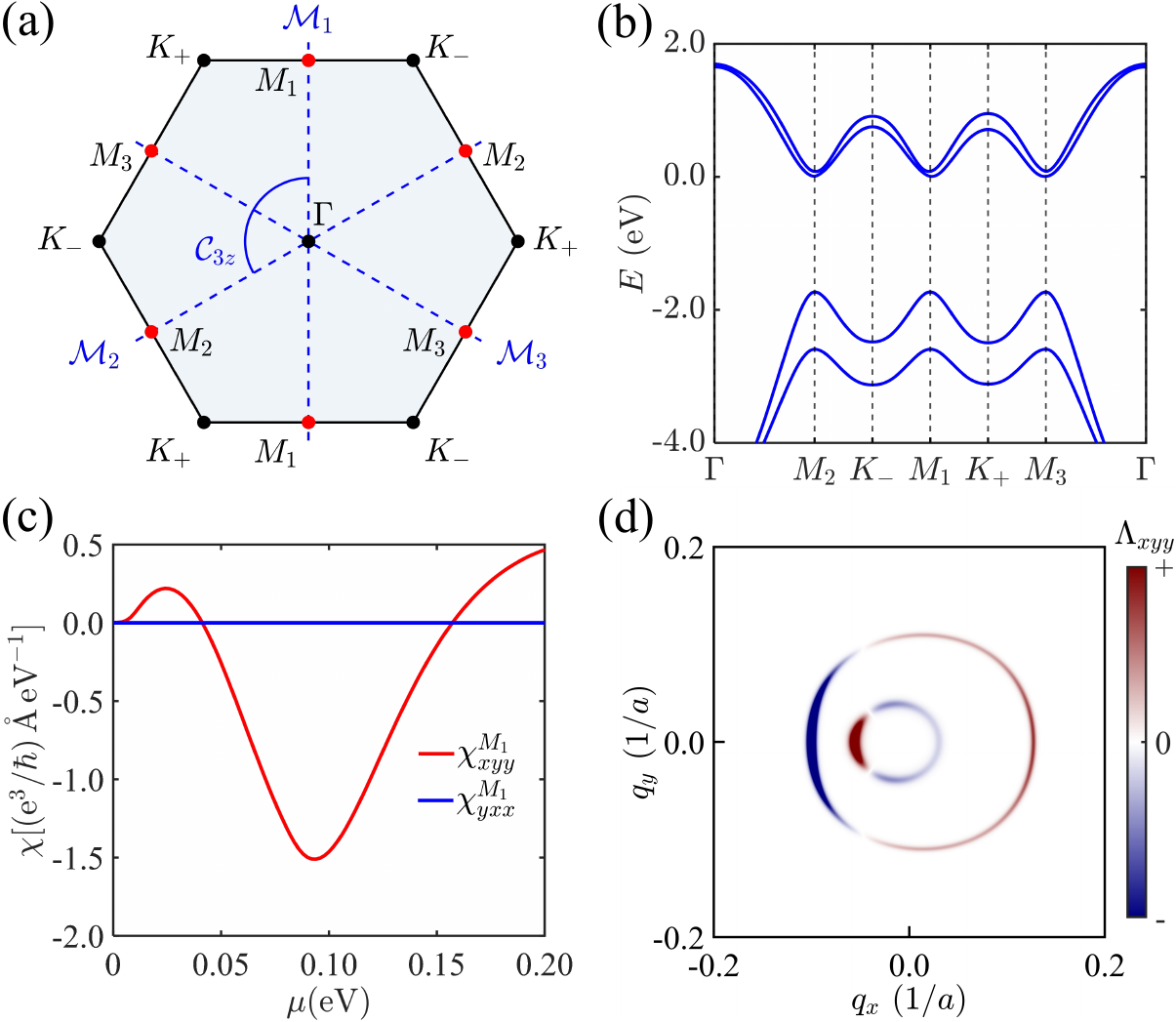}
		\par\end{centering}
	\caption{\label{fig1}Band structure and nonlinear Hall conductivities from $M$ valleys for a tight-binding model. (a) The Brillouin zone with triple $M$ points and their symmetries. (b) Band structure with degenerate $M$ valleys. (c) Conductivity tensor elements $\chi^{M_1}_{xyy}$ and $\chi^{M_1}_{yxx}$ as functions of the chemical potential $\mu$ near the conduction band edge. (d) BCP dipole $\Lambda_{xyy}$ in the vicinity of $M_1$ valley at $\mu = 0.1 \,\mathrm{eV}$, with $q_{x,y}$ being wave vectors with respect to the $M_1$ point. In (b) and (c), the conduction band minima are set to zero.}
\end{figure}

Figure~\ref{fig1}(a) shows the first Brillouin zone (BZ) of the honeycomb lattice, with high-symmetry momentum points.
It is seen that $K$ points are located at the corners of the hexagonal BZ, and $M$ points reside at the midpoints of the sides. 
The $M$ points at opposite sides can be connected by reciprocal lattice vectors and are thus equivalent to each other. 
Accordingly, there are three inequivalent $M$ points in the BZ, which are labeled as $M_{1\text{--}3}$.
While valley-contrasting optoelectronic properties at $K$ valleys have been extensively studied, triple $M$ valleys remain largely unexplored.
We will focus on physical properties of triple $M$ valleys in the following.
Different from $K$ points, $M$ points are invariant under both time reversal $\mathcal{T}$ and spatial inversion $\mathcal{P}$ for a centrosymmetric nonmagnetic lattice.
The presence of either $\mathcal{P}$ or $\mathcal{T}$ symmetry excludes the QM-induced nonlinear Hall responses from $M$ valleys~\cite{Liu2021}.
An out-of-plane N\'{e}el-type antiferromagnetic order is further introduced into the honeycomb lattice to break $\mathcal{P}$ and $\mathcal{T}$ symmetries, thereby enabling the QM-induced nonlinear Hall responses.
Another merit of the magnetic order is that it preserves the $\mathcal{PT}$ symmetry, which suppresses linear and nonlinear Hall responses from Berry curvature and its dipole, thereby removing their interference.
In this magnetic configuration, each $M$ point preserves a local $\mathcal{M}\mathcal{T}$ symmetry, with the vertical mirror $\mathcal{M}$ containing the $M$ point and its normal along the corresponding edge of the BZ.
The three mirrors are denoted as $\mathcal{M}_{1\text{--}3}$, as illustrated in Fig.~\ref{fig1}(a).
The mirror $\mathcal{M}_{i}$ $(i=1,2,3)$ leaves the $M_i$ point invariant, and it also connects the other two $M$ points.
In addition, three $M$ points are related by the threefold rotation symmetry $\mathcal{C}_{3z}$.

Given the honeycomb lattice and $\mathcal{PT}$-symmetric N\'{e}el-type antiferromagnetic order, the tight-binding model takes into account both the sublattice and spin degrees of freedom.
Each sublattice adopts two atomic orbitals, $d_{xz}$ and $d_{yz}$.
This model incorporates nearest-neighbor and next-nearest-neighbor hoppings, on-site atomic spin-orbit coupling and out-of-plane antiferromagnetic exchange interactions.
Its explicit form and associated parameters are given in the Supplemental Material (SM).

Figure~\ref{fig1}(b) displays the tight-binding band structure of the antiferromagnetic honeycomb lattice.
It is found that there are two conduction and two valence bands at the chosen energy windows.
All bands remain spin-degenerate at every wave vector $\boldsymbol{k}$, as protected by $\mathcal{PT}$ symmetry~\cite{Li2013}.
Moreover, both the conduction band minima and the valence band maxima are located in the vicinity of $M_{1-3}$ points.
The electronic states at the three $M$ points are also degenerate, which is ensured by the $\mathcal{C}_{3z}$ symmetry.
Therefore, the band-edge regions surrounding the $M$ points form three inequivalent, degenerate valleys.
Furthermore, compared with the valence bands, the conduction bands exhibit a much smaller splitting, thereby possibly leading to a significant nonlinear electric response, which will become clear shortly.

\textit{\color{blue}QM-induced nonlinear Hall conductivities.} 
Based on the tight-binding model of the centrosymmetric antiferromagnet, we further investigate QM-induced nonlinear Hall currents from $M$ valleys.
For a certain $M_i$ valley, the nonlinear current density, $j_a^{M_i}$, is given by $j_a^{M_i} = \chi_{abc}^{M_i} E_b E_c$, where $\chi_{abc}^{M_i}$ is the nonlinear conductivity tensor element of $M_i$ valley and $E_{b,c}$ are electric-field components, with $a, b, c$ denoting Cartesian coordinates.
For the in-plane Hall transport measurements, Cartesian coordinates are fixed in the $x$-$y$ plane, and accordingly $\chi_{xyy}^{M_i}$ and $\chi_{yxx}^{M_i}$ are the only two independent nonlinear Hall conductivity components, which are studied in what follows.
These tensor elements are expressed as the integrals of BCP dipole, which includes the contribution from band-resolved QM and represents an intrinsic quantum geometric property~\cite{Wang2021,Liu2021}. 
The expressions are provided in SM.

Figure~\ref{fig1}(c) shows $\chi_{xyy}$ and $\chi_{yxx}$ from $M_1$ valley as functions of the chemical potential $\mu$ near the conduction band edges. 
It is seen that $\chi_{xyy}$ varies with $\mu$, while $\chi_{yxx}$ is always zero. 
As $\mu$ increases, $\chi_{xyy}$ first exhibits a small positive peak at $\mu = 24~\mathrm{meV}$, then reverses sign to develop a large negative peak at $\mu = 94~\mathrm{meV}$, and eventually recovers to positive values at higher $\mu$.
Notably, the value of the negative peak is $-1.5 \, (e^3/\hbar) \, \text{\AA} \, \mathrm{eV}^{-1}$, which is comparable to the nonlinear Hall conductivity from $K$ valleys of strained graphene~\cite{Das2024}.
Besides, $\chi_{xyy}$ is calculated near the valence band edge, and its values are substantially smaller than those near the conduction band edge by orders of magnitude.
The considerable $\chi_{xyy}$ near the conduction band edge arises from the smaller band splitting, since the nonlinear response is highly sensitive to the interband energy difference~\cite{Du2018,Zhou2025}.
 
To gain insight into the nonlinear Hall conductivity from $M_1$ valley, the BCP dipole $\Lambda_{xyy}(\boldsymbol{k})$, i.e., the momentum-resolved conductivity, is presented around this valley in Fig.~\ref{fig1}(d).
It is found that $\Lambda_{xyy}(\boldsymbol{k})$ is predominantly concentrated in the vicinity of $M_1$ valley, confirming that this valley, i.e., the band-edge states, dominates the response.
Moreover, $\Lambda_{xyy}(\boldsymbol{k})$ is symmetric with respect to the line $q_y = 0$, which is ensured by the aforementioned $\mathcal{M}_1\mathcal{T}$ symmetry (see SM).
As a result, integrating $\Lambda_{xyy}(\boldsymbol{k})$ over $M_1$ valley gives rise to a finite $\chi_{xyy}^{M_1}$.
In contrast, an antisymmetric $\Lambda_{yxx}(\boldsymbol{k})$ leads to a vanishing $\chi_{yxx}^{M_1}$, as discussed in SM.

Taking one step further from $M_1$ valley, the nonlinear Hall conductivities from $M_2$ and $M_3$ valleys are derived by the $\mathcal{C}_{3z}$ symmetry that relates the three $M$ valleys.
The nonlinear Hall conductivity tensor elements from different $M$ valleys satisfy,
\begin{equation}\label{H6}
	\begin{aligned}
		\chi^{M_{j+1}}_{a'b'c'} = R_{a'a} R_{b'b} R_{c'c} \chi^{M_j}_{abc}, \quad (j = 1, 2)
	\end{aligned}
\end{equation}
where $R$ is the matrix representation of the $\mathcal{C}_{3z}$ operation.
Specific to $\chi_{xyy}$ and $\chi_{yxx}$, Eq.~(\ref{H6}) yields the relations as follows,
\begin{align}
	\chi^{M_{j+1}}_{xyy} &= - \frac{1}{2} \chi^{M_j}_{xyy} - \frac{\sqrt{3}}{2} \chi^{M_j}_{yxx},
	\label{H7} \\[4pt]
	\chi^{M_{j+1}}_{yxx} &= \frac{\sqrt{3}}{2} \chi^{M_j}_{xyy} - \frac{1}{2} \chi^{M_j}_{yxx},
	\label{H8}
\end{align}
which has also been confirmed by numerical calculations.

\begin{figure}[t]
	\begin{centering}
		\includegraphics[width=8.4cm]{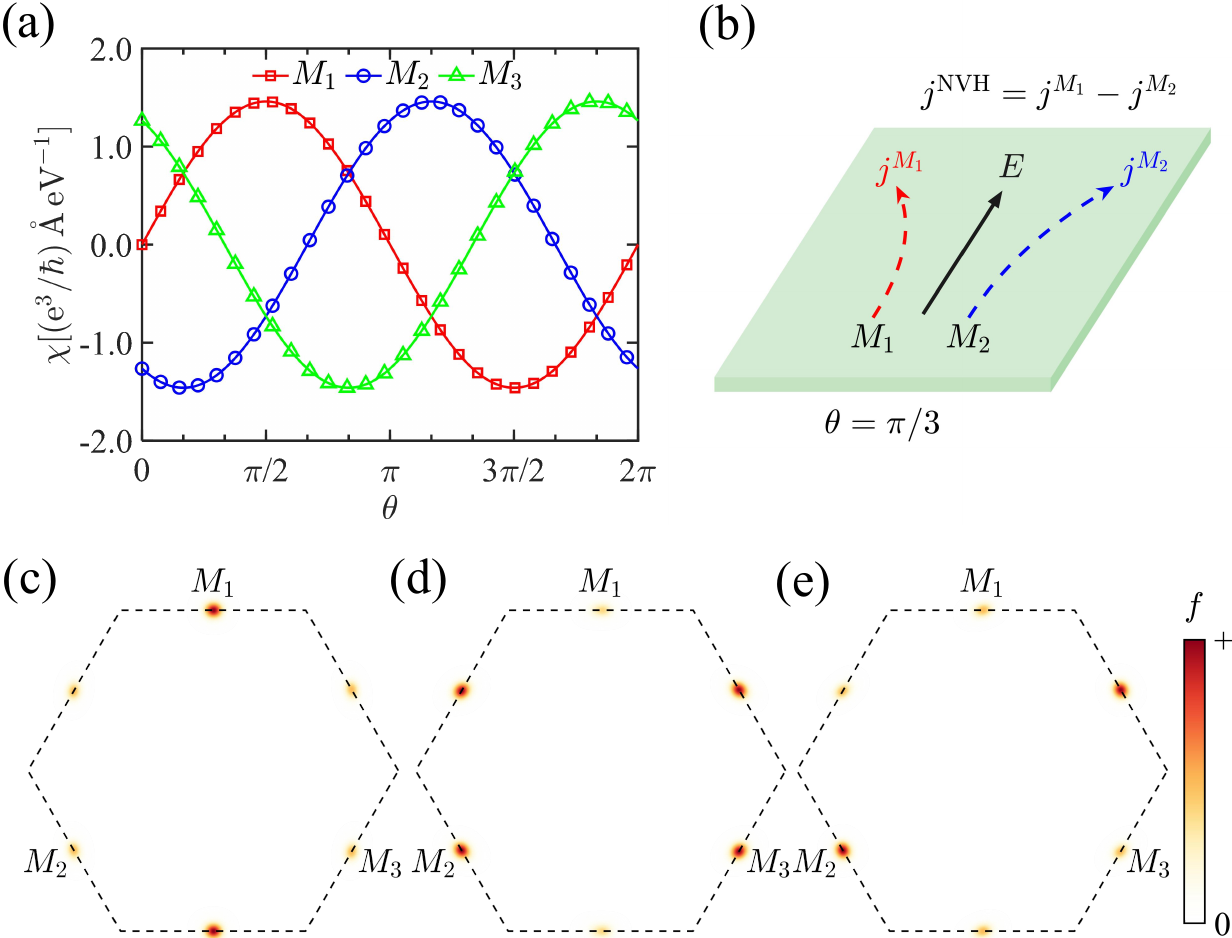}
		\par\end{centering}
	\caption{\label{fig2} High tunability of triple $M$ valleys by electric and optical fields. (a) Evolutions of nonlinear Hall conductivities of $M$ valleys with the electric-field angle $\theta$ at $\mu = 0.1\,\text{eV}$. (b) Schematic of the NVH effect from $M_1$ and $M_2$ valleys. (c)-(e) Momentum-resolved oscillator strength $f(\boldsymbol{k})$ for linearly polarized light with the polarization along the $x$-axis, the $y$-axis, and at an angle of $2\pi/3$ with respect to the $x$-axis. The Brillouin zone is bounded by dashed lines. The photon energy and smearing parameter are set to $1.7\,\text{eV}$ and $0.05\,\text{eV}$, respectively.}
\end{figure}

\textit{\color{blue}Electric-field-tunable NVH effect.} 
An in-plane electric field in an arbitrary direction is further considered to explore NVH conductivities from $M$ valleys.
When the applied electric field is defined as $\boldsymbol{E} = E(\cos\theta, \sin\theta)$ with $\theta$ being the polar angle with respect to the $x$-axis,
the nonlinear Hall conductivity for $M_i$ valley is given as
$\chi^{M_i}_{\mathrm{H}}(\theta) = \chi_{yxx}^{M_i} \cos\theta - \chi_{xyy}^{M_i} \sin\theta$~\cite{Liu2021}.
Combining the vanishing $\chi_{yxx}^{M_1}$ in Fig.~\ref{fig1}(c) and the Eqs.~\ref{H7}--\ref{H8}, the nonlinear Hall conductivities of three $M$ valleys for an electric field in an arbitrary direction are explicitly expressed as
\begin{equation}\label{H9}
	\chi^{M_1}_{\mathrm{H}} = -\chi_{xyy}^{M_1} \sin\theta,\quad
	\chi^{M_{2,3}}_{\mathrm{H}} = \chi_{xyy}^{M_1} \sin\left( \theta \pm \frac{\pi}{3} \right).
\end{equation}

According to Eq.~\ref{H9}, Fig.~\ref{fig2}(a) presents nonlinear Hall conductivities of three $M$ valleys at $\mu = 0.1~\text{eV}$. 
The Hall conductivities all exhibit $2\pi$-periodic sinusoidal dependences on the angle $\theta$, with a relative phase shift of $2\pi/3$. 
Consequently, for each $M$ valley, both the magnitude and sign of the Hall conductivity can be continuously tuned by rotating the electric field. 
At $\theta = l\pi/3$ ($l = 0, 1, \dots, 6$), the Hall conductivity of a certain $M$ valley vanishes completely, while those of the other two valleys have the same magnitude but opposite signs.
Taking $\theta = \pi/3$ as an example, it is found that $j^{M_3} = 0$ and $j^{M_1} = -j^{M_2}$.
Therefore, the nonlinear Hall currents from $M_1$ and $M_2$ valleys flow in opposite directions, leading to a NVH current, $j^{\mathrm{NVH}} = j^{M_1} - j^{M_2}$, as depicted in Fig.~\ref{fig2}(b).
Moreover, the total charge Hall current always vanishes under the protection of the $\mathcal{C}_{3z}$ symmetry, so that only the pure NVH current remains.
Besides, for a general electric-field direction with $\theta \neq l\pi/3$, the net NVH current can also be induced but with the contributions from all three $M$ valleys.

Taken together, the electric-field direction determines which pair of $M$ valleys generates the net NVH current, offering a practical route to electrically manipulate NVH transport with the full harnessing of three flavors of $M$ valleys.
It is also noted that the electric-field-dependent valley Hall effect is a prominent signature of $M$ valleys, which is absent at $K_\pm$ valleys.
The local $\mathcal{C}_{3z}$ symmetry of $K_\pm$ valleys enforces a constant valley Hall current regardless of the electric-field direction. 

\textit{\color{blue} QM-based optical excitation.} 
Beyond nonlinear Hall transport, the QM also governs the optical selection rule for linearly polarized light~\cite{Li2026Optical}.
We further investigate the optical selectivity in triple $M$ valleys of the $\mathcal{PT}$-symmetric antiferromagnet, which may be used for selective photoexcitation of $M$ valleys.
The calculation details are given in SM.

Figures~\ref{fig2}(c)--(e) demonstrate the total momentum-resolved oscillator strength, $f(\boldsymbol{k})$, for linearly polarized optical transitions between the valence and conduction bands, with the polarization oriented along the $x$-axis, the $y$-axis and at an angle of $2\pi/3$ with respect to the $x$-axis, respectively, at the photon energy matching the band gap.
For the $x$-polarized light, $f(\boldsymbol{k})$ has the maximal value at $M_1$ valley, while it is smaller at $M_2$ and $M_3$ valleys, indicating that the linearly polarized light predominantly excites carriers from $M_1$ valley.
As for the $y$-polarized light, the relative magnitudes of $f(\boldsymbol{k})$ reverse among these valleys, with $M_1$ valley exhibiting the minimal value, and carriers are thus mainly excited from $M_2$ and $M_3$ valleys.
The variation of $f(\boldsymbol{k})$ for $M_1$ valley arises from the constraint of the $\mathcal{M}_1\mathcal{T}$ symmetry, which guarantees that the extrema appear for light polarized parallel or perpendicular to the mirror plane~\cite{Li2026Optical}. 
Moreover, the maximum of $f(\boldsymbol{k})$ shifts to the $M_2$ valley when the light has a polarization angle of $2\pi/3$, as shown in Fig.~\ref{fig2}(e), and the maximum moves to $M_3$ valley at a polarization angle of $4\pi/3$.
Therefore, all three $M$ valleys can be preferentially excited by linearly polarized light.

By combining the QM-induced optical selectivity and nonlinear Hall current, a valley-polarized Hall effect can be generated with a net anomalous Hall current.
Specifically, compared with the NVH effect in Fig.~\ref{fig2}(b) featuring two valley currents with identical magnitudes, applying linearly-polarized light along the $x$-axis (at a polarization angle of $2\pi/3$) further increases the carrier population in $M_1$ ($M_2$) valley, which drives a valley-polarized Hall current dominated by this valley. 
Therefore, $M$ valleys can be optically selected and electrically detected, enabling multistate information manipulation by virtue of their three flavors.

\begin{figure}[t]
	\begin{centering}
		\includegraphics[width=8.4cm]{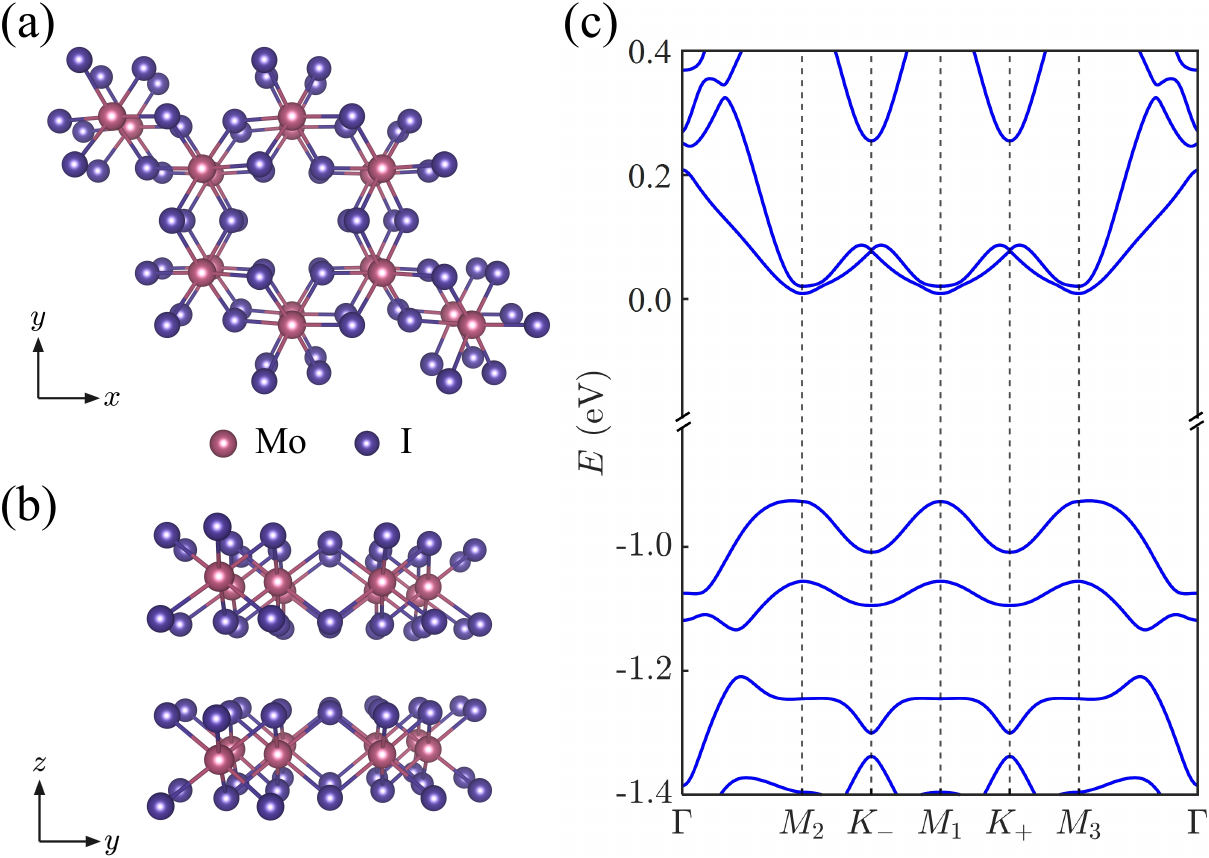}
		\par\end{centering}
	\caption{\label{fig3}Crystal and band structures of the $\mathrm{MoI}_3$ bilayer. (a) Top and (b) side views of the crystal structure. (c) Band structure with the conduction band minima set to zero.}
\end{figure}

\textit{\color{blue} Material realization.} 
To realize the above optoelectronic responses from $M$ valleys in a concrete material, we consider a $\mathcal{P}\mathcal{T}$-symmetric AA-stacked $\mathrm{MoI}_3$ bilayer as a promising candidate~\cite{Haastrup2018}.
Figures~\ref{fig3}(a) and \ref{fig3}(b) show top and side views of the bilayer.
For a $\mathrm{MoI}_3$ monolayer, Mo atoms form a planar honeycomb lattice, and each Mo atom is octahedrally coordinated by surrounding I atoms.
When one monolayer is directly placed on the other without any lateral shift, the AA-stacked bilayer is constructed with the space group $P\overline{3}1m$ (No.~162).
Given that the $\mathrm{MoI}_3$ monolayer is an intrinsic out-of-plane ferromagnetic semiconductor, an interlayer antiferromagnetic coupling endows the bilayer with the $\mathcal{P}\mathcal{T}$ symmetry.
Besides, the $\mathcal{C}_{3z}$ and $\mathcal{M}_y\mathcal{T}$ symmetries are also present in the bilayer.
By first-principles calculations, the in-plane lattice constant of the bilayer is $7.07\,\text{\AA}$ and the magnetic moment is dominantly contributed by Mo atoms, with a value of $2.4\,\mu_{\mathrm{B}}$ per Mo atom.
The interlayer antiferromagnetism is verified to be more stable than the interlayer ferromagnetism by $14.2\,\text{meV}$.

Figure~\ref{fig3}(c) displays the band structure of the $\mathrm{MoI}_3$ bilayer with the out-of-plane magnetization.
All bands are spin degenerate, and the band extrema are located in the vicinity of three $M$ points for both the valence and conduction bands, yielding three degenerate $M$ valleys.
Besides, a small band splitting appears between the conduction bands.
The first-principles results are similar with the tight-binding model.

\begin{figure}[t]
	\centering
	\includegraphics[width=8.4cm]{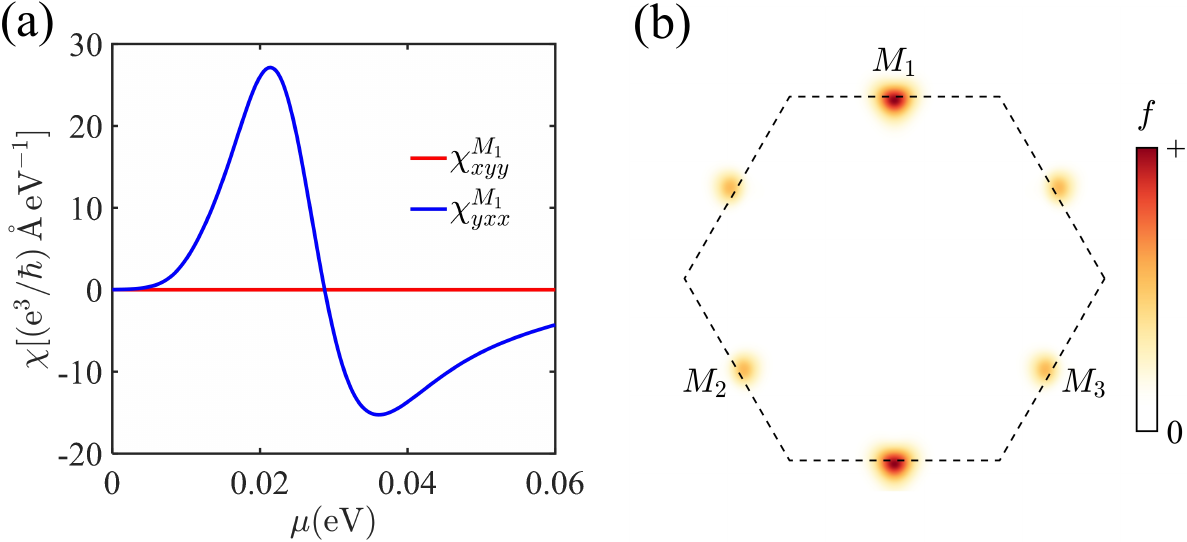}
	\caption{\label{fig4}QM-induced optoelectronic responses in the $\mathrm{MoI}_3$ bilayer. (a) Conductivity tensor elements $\chi^{M_1}_{xyy}$ and $\chi^{M_1}_{yxx}$ as functions of $\mu$. (b) Oscillator strength $f(\boldsymbol{k})$ for the optical polarization along the $x$-axis. The photon energy and smearing parameter are set to $0.93\,\text{eV}$ and $0.01\,\text{eV}$, respectively.}
\end{figure}

Figure~\ref{fig4}(a) presents evolutions of the conductivity tensor elements $\chi^{M_1}_{xyy}$ and $\chi^{M_1}_{yxx}$ from $M_1$ valley with the chemical potential $\mu$ near the conduction band edge.
It is seen that $\chi^{M_1}_{yxx}$ varies with $\mu$, whereas $\chi^{M_1}_{xyy}$ vanishes.
The nonzero and zero components are thus interchanged compared with the model results in Fig.~\ref{fig1}(c).
This is because the relevant symmetry of the realistic material is $\mathcal{M}_y\mathcal{T}$ instead of $\mathcal{M}_x\mathcal{T}$ (i.e., $\mathcal{M}_1\mathcal{T}$) in the model.
With increasing $\mu$, $\chi^{M_1}_{yxx}$ exhibits a positive peak of $27.1\,(e^3/\hbar)\,\text{\AA}\,\text{eV}^{-1}$ at $\mu = 21.5\,\text{meV}$, followed by a negative peak of $15.3\,(e^3/\hbar)\,\text{\AA}\,\text{eV}^{-1}$ at $\mu = 36.1\,\text{meV}$.
The peak magnitudes surpass the nonlinear Hall conductivity experimentally observed in CuMnAs films by one order of magnitude~\cite{Godinho2018,Wang2023} and they are thus expected to be detectable via e.g., nonlocal resistance measurements~\cite{Sui2015gate,Shimazaki2015,Cao2025Nonlocal}.
Moreover, the optical selectivity is also confirmed in the 
$\mathrm{MoI}_3$ bilayer.
In Fig.~\ref{fig4}(b), the $x$-polarized light is found to mainly excite the $M_1$ valley.
Further considering the tunabilities from electric and optical fields, QM-induced optoelectronic responses enable three $M$ valleys to be distinguished in the bilayer.

\textit{\color{blue} Discussion and summary.} 
In the above tight-binding and first-principles calculations, the magnetization is taken along the $z$-axis, which preserves the $\mathcal{C}_{3z}$ symmetry and valley degeneracy.
The $\mathcal{C}_{3z}$ symmetry can be further broken by an in-plane magnetization, which may provide an additional pathway to tailor the nonlinear transport.
The QM-induced nonlinear Hall conductivities in the antiferromagnetic $\mathrm{MoI}_3$ bilayer with the in-plane magnetization are provided in the SM.
It is found that the in-plane magnetization allows a nonlinear valley-polarized Hall current.
Specifically, nonlinear Hall currents with different magnitudes from three $M$ valleys may flow in the same or opposite directions, depending on the in-plane magnetization orientation or the chemical potential, and they cannot cancel out, resulting in a net charge current.
The in-plane magnetization thus further offers an experimentally accessible knob to modulate the nonlinear Hall transport from $M$ valleys.
Besides, external stimuli, e.g., uniaxial strain, can also applied to break the $\mathcal{C}_{3z}$ symmetry and tune optoelectronic responses.

It is also noted that the $\mathrm{MoI}_3$ bilayer studied above constitutes merely one material realization of such QM-induced optoelectronic responses from triple $M$ valleys.
These responses can be generalized to a broad class of $\mathcal{PT}$-symmetric antiferromagnets hosting $M$ valleys.
For example, the $\mathrm{MnTe}$~\cite{Sattar2022} and $\mathrm{CoSe}$~\cite{Li2026CoSe} monolayers with intralayer antiferromagnetism, as well as the $\mathrm{LaBr}_{2}$ bilayer with interlayer antiferromagnetism~\cite{Chen2022}, are also highly promising.
Furthermore, even in centrosymmetric nonmagnetic materials with $M$ valleys~\cite{Yu2024,Calugaru2025,Lei2025}, the nonlinear Hall current is likely to be utilized for selectively expressing triple $M$ valleys.
For the nonmagnetic materials, while both linear and second-order responses to the electric field are symmetry-forbidden, the higher-order response is expected to arise from $M$ valleys~\cite{Liu2022third,Liu2025,Gong2026}.
Besides, other inequivalent valleys related by the $\mathcal{C}_{3z}$ symmetry, e.g., $Q$ valleys in transition-metal dichalcogenides~\cite{Wu2016}, may well also exhibit QM-induced optoelectronic effects.
The above systems merit further investigation.

In summary, our work reveals that QM-induced optoelectronic responses enable the recognition and modulation of three flavors of $M$ valleys. 
Nonlinear Hall currents from $M$ valleys are highly tunable.
They can be generated from any pair of valleys or predominantly from an individual valley, by virtue of electric and optical fields. 
Our findings unlock the potential of triple $M$ valleys for valleytronic applications and provide opportunities for multistate information encoding and manipulation.

\textit{\color{blue}\textit{Acknowledgments}} -- We are supported by the National Natural Science Foundation of China (Nos. 12374044, 11904173, 12004186).

\end{document}